\documentclass[preprint,prd]{revtex4-1}
\usepackage{color}
\usepackage{graphicx}
\usepackage{epsfig}
\usepackage{subfig}

\usepackage{graphicx}
\usepackage{dcolumn}
\usepackage{bm}

\begin{document}

\title{Non-minimal  $\ln(R)F^2$  Couplings of Electromagnetic 
Fields to Gravity: Static, Spherically Symmetric  Solutions }

\author{Tekin Dereli}
\email{tdereli@ku.edu.tr}
 \affiliation{Department of Physics,
Ko\c{c}
University, 34450 Sar{\i}yer, \.{I}stanbul, Turkey}

\author{\"{O}zcan Sert}%
 \email{sertoz@itu.edu.tr}

\affiliation{Department of Physics Engineering\\ \.{I}stanbul
Technical University, 34469 Maslak, \.{I}stanbul, Turkey
}%


\date{\today}

\begin{abstract}

 \noindent 
 We   investigate    the  non-minimal  couplings between the electromagnetic fields and gravity through the natural logarithm of the curvature scalar.   After we give the Lagrangian formulation of the non-minimally    coupled 
theory, we derive 
field equations  by a first order variational principle using the
method of Lagrange multipliers. We look at  static,
spherically symmetric solutions that are asymptotically flat. We discuss the nature of horizons for some candidate black hole solutions  according to  various values of  the parameters $R_0 $ and $a_1$.



\end{abstract}

\pacs{Valid PACS appear here}
\maketitle


\def\ba{\begin{eqnarray}}
\def\ea{\end{eqnarray}}
\def\w{\wedge}



\section{Introduction}

\noindent  

 We consider
the non-minimal couplings of gravitational and electromagnetic fields
 described by a Lagrangian density.  Such couplings may  occur  near compact astrophysical objects which has  high mass density such as the neutron stars or black holes.
The  non-minimally  coupled   electromagnetic fields to gravity  in $RF^2$ form 
were extended  and classified  \cite{prasanna,horndeski} to gain
more insight into the relationship between  space-time curvature 
and charge conservation.
They  were also  obtained  from a calculation   in QED  of  the photon effective action from 1-loop vacuum polarization
on a curved background \cite{drummond} and from Kaluza-Klein reduction  of a five-dimensional
$R^2$-Lagrangian \cite{buchdahl,dereli1}.
 A non-minimally coupled Einstein-Maxwell Lagrangian in general may    involve  field equations of
order higher than two. However,   the type of non-minimal couplings we consider are obtained   by the   reduction of the Euler-Poincar\'{e} Lagrangian  
in five dimensions to four dimensions and involve at most second order field equations \cite{muller-hoissen}.
   A three parameter family of
non-minimally coupled Einstein-Maxwell  field equations  was
studied in  \cite{balakin1}.

Recently, in the context of  primordial magnetic fields present  during the reheating epoch of the universe,  the $R^mF^2$-type couplings were discussed in \cite{lambiase}. The modified gravity with $lnR$ terms  \cite{nojiri} has also got much attention, since it could  explain the observed acceleration of  the universe  and  the dark energy concept without using any exotic fields.
  The behavior of the rotational velocities of test particles gravitating around galaxies is investigated in modified gravity in the context of non-minimal matter couplings \cite{harko1,harko2}.   In this context, 
    the modified  $f(R)$-Maxwell  
gravity with  $I(R) F^2$ coupling terms   \cite{bamba1, bamba2} and $f(G)$-Maxwell gravity in which non-minimal coupling between electromagnetic fields and 
a function of  Gauss-Bonnet invariant  \cite{setare} were proposed  
to explain late-time cosmic acceleration.
 There are many other theories of gravity with non-minimal couplings \cite{nojiri2}, \cite{nojiri3}. However, 
 the non-minimal couplings with electromagnetic fields are not  investigated in sufficient detail. Especially, finding spherically symmetric solutions is not an easy task for such theories \cite{balakin2},\cite{balakin3}. Furthermore, any arbitrary non-minimal coupling may not give rise to  solutions satisfying physical asymptotic conditions and observations in solar and cosmological scales. Therefore,  here we propose a  non-minimal theory with a special  $lnR$ coupling term.

We first discuss   a non-minimal $Y(R)F^2$ coupled
Einstein-Maxwell theory    using the
algebra of exterior differential forms.   We derive the field
equations by a first order variational principle using the method
of Lagrange multipliers.  We choose in  particular $Y(R)=\frac{1}{1-a1ln{R/R_0}}$ and look for static, spherically symmetric solutions that are asymptotically flat. 
 In our model, the $ Y(R) F^2$ coupled term  in the Lagrangian leads to modifications   both in the Maxwell and Einstein field equations. 
The modifications in the Maxwell equations can  be related with the  polarization and the  magnetization   in a specific medium.    The non-minimal couplings also give  important modifications   to the structure of a charged 
black hole. These may shed light on some  important  problems of gravity such as dark matter and dark energy without  introducing a cosmological constant or any other type of scalar fields.

\section{Field Equations of the Non-minimally Coupled Theory} \label{model}

\bigskip

\noindent We will derive our field equations by a variational
principle from an action
\begin{equation}
        I[e^a,{\omega^a}_b,F] = \int_M{L} = \int_M{\mathcal{L}^*1},
        \nonumber
\end{equation}
where  $\{e^a\}$ and ${\{\omega^a}_b\}$ are the fundamental
gravitational field variables and   $F$ is the electromagnetic
field 2-form.  The space-time metric $g = \eta_{ab} e^a \otimes
e^b$ with signature $(-+++)$ and we fix the orientation by setting
$*1 = e^0 \w e^1 \w  e^2 \w e^3 $.  Torsion 2-forms $T^a$ and
curvature 2-forms $R^{a}_{\; \; b}$ of spacetime are found from
the Cartan-Maurer structure equations
\begin{equation}
T^a = de^a + \omega^{a}_{\;\;b} \w e^b , \nonumber
\end{equation}
\begin{equation}
R^{a}_{\;\;b} = d\omega^{a}_{\;\;b} + \omega^{a}_{\;\;c} \w \omega^{c}_{\;\;b} . \nonumber
\end{equation}
We consider  the following  Lagrangian density 4-form;
 \ba\label{lag1}
  L =  \frac{1}{2\kappa^2} R*1 -\frac{1}{2}Y(R) F\w *F,
   \label{Lagrange}
   \ea
where
 $\kappa^2 = 8\pi G$ is  Newton's universal gravitational constant $(c=1)$ and $R$ is the curvature scalar which can be found by applying interior product $\iota_a $ twice to the curvature tensor $R_{ab}$ 2-form. We use the shorthand notation $ e^a \wedge e^b \wedge \cdots =
e^{ab\cdots}$, and  $\iota_aF =F_a, \  \  \iota_{ba} F =F_{ab}, $ \   $ \iota_a {R^a}_b =R_b, \  \   \iota_{ba} R^{ab}= R $. 
   The field equations are obtained by considering the independent variations of
   the action with respect to  $\{e^a\}$,
   ${\{\omega^a}_b\}$ and $\{F\}$.  The electromagnetic field components are read  from the expansion $F = \frac{1}{2} F_{ab} e^a \w e^b$.
We will confine ourselves to the unique metric-compatible, torsion-free
Levi-Civita connection. We impose this choice of connection
through the  constrained variations of the action  by the method of Lagrange
multipliers. That is, we add to the above Lagrangian density  the
following constraint terms:
\begin{equation}
L_{C} = \left ( de^a + \omega^{a}_{\;\;b} \w e^b \right ) \w
\lambda_a + dF  \w \mu  \nonumber,
\end{equation}
where   $\lambda_a$'s are  Lagrange multiplier 2-forms whose
variation imposes the zero-torsion constraint  $T^a=0$. We also
use a first order variational principle for the electromagnetic
field 2-form $F$ for which the homogeneous field equation $dF = 0$
is imposed by the variation of the Lagrange multiplier 2-form
$\mu$.

\medskip

\noindent The infinitesimal variations of the total Lagrangian
density $L + L_C$ (modulo a closed form) are given by
\begin{eqnarray}\label{generaleinsteinfe1}
&& \dot{L} +{\dot{L}_C} = \frac{1}{2 \kappa^2} \dot{e}^a \w R^{bc}
\w *e_{abc} +  \dot{e}^a \w \frac{1}{2} Y(R)   (\iota_a F \w *F - F \w \iota_a *F)   +  \dot{e}^a \w D \lambda_a \nonumber
\\ & &
 + \dot{e}^a \w     Y_R     (\iota_a R^b)(\iota_b F \w *F  + F \w \iota_b *F)   + \frac{1}{2} \dot{\omega}_{ab} \w  ( e^b
\w \lambda^a - e^a \w \lambda^b)
  \nonumber \\
& &
 \dot{\omega}_{ab} \w  {\Sigma}^{ab} 
 -\dot{ F} \w Y(R)   *F   +
\dot{\lambda}_a \w T^a   - \dot{F} \w d\mu .
\end{eqnarray}
where  $Y_R = \frac{dY}{dR}$, and the angular momentum tensor
\begin{eqnarray}\label{sigmaab1}
 {\Sigma}^{ab} &=& - \frac{ 1}{2} D  [ Y_R(  F^{ab} *F 
  +  F^b \wedge \imath^a * F-  F^a \wedge \imath^b* F 
-   F\wedge \imath^{ab}*F) ].
   \end{eqnarray}
The Lagrange multiplier 2-forms $\lambda_a$ are solved  uniquely
from the connection variation equations 
 \begin{eqnarray}\label{lambdaaeb}
 e_a\w \lambda_b -  e_b \w \lambda_a = 2{\Sigma}_{ab},
 \end{eqnarray}
by applying the  interior product operator twice as
\begin{eqnarray}\label{lambdaaeb2}
\lambda^a &=&  2\imath_b   {\Sigma}^{ba}  +\frac{1}{2} 
\imath_{bc}  {\Sigma}^{cb}\wedge e^a.
\end{eqnarray}
We substitute the $ \lambda_a$\rq{}s into the $\dot{e^a}$ equations and after some simplifications we find the  Einstein field equations for the extended theory as
\begin{eqnarray}\label{einstein}
  \frac{1}{2 \kappa^2}  R^{bc}
\w *e_{abc} +  \frac{1}{2} Y  (\iota_a F \w *F - F \w \iota_a *F)  &&  + Y_R  (\iota_a R^b)\iota_b( F \w *F )
 \nonumber
\\ & &
+ \frac{1}{2}  D [ \iota^b D(Y_R F_{mn} F^{mn} )]\wedge *e_{ab}
 =0 ,
\end{eqnarray}
while the Maxwell\rq{}s equations read
\begin{equation}\label{maxwell1}
dF = 0 \quad , \quad  d* (Y   F) = 0 .
\end{equation}
In terms of a local inertial coordinate system $(x^\mu)$,
the master equations (\ref{einstein}) and (\ref{maxwell1}) turn out to be equivalent to
the following equations  that  were obtained  by Bamba and Odintsov \cite{bamba2}  by a variation procedure with respect to the metric:
\begin{eqnarray}
R_{\mu \nu} -\frac{1}{2}g_{\mu \nu} R =\kappa^2T_{\mu \nu},
\end{eqnarray}
where the electromagnetic energy-momentum tensor components 
\begin{eqnarray}
T_{\mu \nu } =&& Y(g^{\alpha \beta} F_{\mu \beta} F_{\nu \alpha} -\frac{1}{4} g_{\mu \nu } F_{\alpha \beta} F^{\alpha \beta}) \nonumber \\
&&+\frac{1}{2}\{ Y_R F_{\alpha \beta} F^{\alpha \beta} R_{\mu \nu} + g_{\mu \nu}g^{\gamma \sigma}  \nabla_\gamma \nabla_\sigma  [Y_R F_{\alpha \beta} F^{\alpha \beta}] -
\nabla_\mu \nabla_\nu [Y_R F_{\alpha \beta} F^{\alpha \beta} ]\},
\end{eqnarray}
and the Maxwell equations
\begin{eqnarray}
 \frac{1}{\sqrt{-g}}\partial_\mu [ \sqrt{-g} \tilde{F}^{\mu \nu}] =0 \quad , \quad  \frac{1}{\sqrt{-g}}\partial_\mu [ \sqrt{-g} Y F^{\mu \nu}] =0  .
\end{eqnarray}

\noindent  The effects of non-minimal
couplings of the electromagnetic fields to gravity  can be expressed  by a constitutive tensor. For this purpose, Maxwell's equations satisfied by
electromagnetic field $F$ in an arbitrary medium can be written as
\ba dF = 0 \quad , \quad *d*G = 0 \ea where $G$ is called the
excitation 2-form. We  have taken  the source-free  Maxwell equations. The effects of gravitation and electromagnetism on matter
are described by $G$.
We can complete this system writing the following  linear constitutive relation \ba G =
{\cal{Z}} (F) \ea where ${\cal{Z}}$ is a type-(2,2)-constitutive
tensor. For the above theory, we have 
\ba G = Y(R) F . \ea 
  We can further introduce the  polarization 1-form $\emph{p} \equiv \emph{d} -
\emph{e} =  (Y-1) \iota_{U} F$  and magnetization
1-form $\emph{ m} \equiv \emph{b} - \emph{h} =   (1-Y) \iota_{U}*F $  relative to a time-like unit velocity vector field U of an inertial observer ( For more details  see \cite{dereli2, dereli3, dereli4}).

 \noindent Now we consider   the following function $Y(R)$ as a special case:
    \ba
  Y(R)=\frac{1}{1-a_1\ln \frac{R}{R_0}} 
   \label{yr}
   \ea
   where  $ a_1  $ is a dimensionless  coupling
   constant and $R_0$ is a constant with the same dimension as that of R.
   We  take $a_1=0$  to get back to the minimal Einstein-Maxwell theory.
   Furthermore, if we look at the limit $R \rightarrow R_0$
 we again obtain the minimal Einstein-Maxwell case.
  It is interesting to note that in the limits   as  R goes either  to zero or  to infinity, the function $Y(R) \rightarrow 0$.  That is, the effects of gravity   become important and  the electromagnetic effects can be neglected in these limits.  On the other hand, as R approaches    $R_0e^{1/a_1}$,   $Y(R) $ becomes very large so that the electromagnetic effects dominate.    Finally,
 the function $Y(R)$ can be expanded as a power series in  $ ln R$  for  $a_1\neq 0$
 and  $|a_1\ln{\frac{R}{R_0}}|<1$   as
   \begin{eqnarray}
\frac{1}{1-a_1\ln \frac{R}{R_0}}=   {\sum\limits_{n=0}^{\infty}}(a_1\ln \frac{R}{R_0})^n.
   \end{eqnarray}
   In this form the model  resembles the RG improved theory  proposed  by Bamba and Odintsov in Ref.\cite{bamba2}.  
   Following this article,   we try to relate the asymptotic freedom in a non-Abelian SU(2)   gauge theory  with a
non-minimal Maxwell-modified gravity by setting  
\begin{eqnarray}
\frac{11\tilde{g}^2 }{12\pi^2}\tilde{t} =   {\sum\limits_{n=1}^{\infty}}(a_1\ln \frac{R}{R_0})^n
   \end{eqnarray}
   where $\tilde{t}$ is a renormalization-group parameter, $\tilde{g}(\tilde{t})$ is the running SU(2) gauge coupling constant and $\tilde{g}=\tilde{g}(0)$. 
   Therefore, if   $|a_1\ln{\frac{R}{R_0}}|<<1$,  we obtain a similar kind of  the RG parameter which has been proposed in \cite{elizalde}.  
    

%
 \section{Static, Spherically Symmetric Solutions }

\noindent We seek static, spherically symmetric  solutions to the field equations which are given by the metric
\begin{equation}\label{metric}
              g = -f(r)^2dt^2  +  f(r)^{-2}dr^2 + r^2d\theta^2 +r^2\sin(\theta)^2 d \phi^2
\end{equation}
and  a static electric potential 1-form  $A =
V(r) dt $. Then
 \begin{eqnarray}\label{electromagnetic}
 F  &=&  dA =V\rq{}dr \w dt   = Edr \w dt .
\end{eqnarray}   

\subsection{Reduced Field Equations}
After a lengthy
calculation we reduce the   non-minimally coupled
Einstein-Maxwell equations  (\ref{einstein}) and (\ref{maxwell1})  for the metric (\ref{metric}) and the electromagnetic 2-form (\ref{electromagnetic}) to the following system of equations:
 \begin{eqnarray}\label{e1}
  && \frac{1}{\kappa^2}(\frac{{f^2}\rq{}}{r}  + \frac{f^2-1 }{r^2} ) - Y_R E^2 (\frac{ {{f^2}\rq{}}\rq{} } {2} + \frac{{f^2}\rq{} }{r}  ) + \frac{1}{2} YE^2
  - [(E^2  Y_R )\rq{} f]\rq{}f  - \frac{2}{r} f^2 (E^2 Y_R   )\rq{}=0,
  \nonumber \\
    &&
\frac{1}{\kappa^2}(\frac{{f^2}\rq{}}{r}  + \frac{f^2-1 }{r^2} ) -  Y_R E^2 (\frac{ {{f^2}\rq{}}\rq{} } {2} + \frac{{f^2}\rq{} }{r}  ) + \frac{1}{2} YE^2
 - (E^2  Y_R )\rq{} ( \frac{{f^2}\rq{}}{2} + \frac{2 f^2}{r} ) =0 ,\\
 && 
 \frac{1}{\kappa^2}(\frac{ {{f^2}\rq{} }\rq{} }{2}  + \frac{{f^2}\rq{}}{r} ) - Y_R E^2 ( \frac{  {{f^2}\rq{}} } {r} + \frac{{f^2-1} }{r^2}  ) - \frac{1}{2} YE^2
  - [(E^2  Y_R )\rq{} f]\rq{}f  - (E^2  Y_R )\rq{} ( \frac{{f^2}\rq{}}{2} + \frac{f^2}{r} )=0, \nonumber
 \end{eqnarray}   
\begin{eqnarray}\label{m2}
Y E=\frac{q}{r^2} .
\end{eqnarray}
Here the curvature scalar
\begin{eqnarray}
R=- {{f^2}\rq{}}\rq{} -\frac{4 }{r} {f^2}\rq{} -\frac{2}{r^2} ( f^2-1) .
\end{eqnarray}
   q is the  electric charge determined by the Gauss  integral 
   \begin{eqnarray}
    \frac{1}{4\pi} {\int_{S^2}{* G }} =  \frac{1}{4\pi} {\int_{S^2}{Y(R) E(r) r^2 \sin \theta  d\theta \wedge d\phi}}=q. 
    \end{eqnarray}
    \subsection{Exact Solution}
At this point, we will make a simplifying assumption and consider solutions that satisfy 
\begin{eqnarray}\label{dif3}
R=\frac{C }{r^4} 
\end{eqnarray}
where $C$ is a constant to be fixed. 
Then,  the field equations (\ref{e1}) and (\ref{m2}) simplify to
\begin{eqnarray}\label{dif4}
\frac{1}{2}\left( {{f^2}\rq{}}\rq{} -\frac{2}{r^2}(f^2-1) \right) \left( \frac{1}{\kappa^2}+ E^2   \frac{a_1}{R(1-a_1\ln
\frac{R}{R_0})^2} \right) -\frac{qE}{r^2}&=&0 , \\
\frac{R}{2}(1 - \frac{a_1\kappa^2E^2}{R(1-a_1 \ln
\frac{R}{R_0})^2} )
&=&0,
\\ \label{dif5}
\frac{1}{1-a_1ln\frac{R}{R_0}}Er^2= q.
\end{eqnarray} 
Thus,   for $a_1 \neq 0$, we can solve the unknown functions for
 the non-minimally extended Einstein-Maxwell theory with $C=a_1 \kappa^2 q^2 $ and  find
\begin{eqnarray}\label{nonminimalsol}
f^2(r) &=& 1-\frac{2M}{r}+\frac{a_1\kappa^2q^2}{r^2}\ln \frac{r}{r_0}  +\frac{\kappa^2q^2 (1+ 5a_1)} {4r^2}
   ,\\
 E(r) &= & \frac{ q}  { r^2}  + \frac{  4q a_1\ln{\frac{r}{r_0} }  }{ r^2}  .
\end{eqnarray}
$r_0$ is an integration constant satisfying the relation $r_0^4=\frac{a_1\kappa^2 q^2}{R_0}$.
The above solution is asymptotically flat.  It is interesting to note that the electric field changes sign (due to polarization effects)  at  $r=r_0 e^{-\frac{1}{4a_1}}$  that corresponds to  
the value $R=R_0 e^{1/a1}$.
On the other hand, the  displacement vector field is an inverse square force field and hence does not change sign.
We observe that  both the curvature   scalar and the   quadratic curvature invariant $*(R_{ab} \wedge * R^{ab})$ for our solutiuon  are singular at the origin $r=0$.  

\subsection{Horizons and Asymptotic Behavior}
 The roots of the metric function $f^2(r)$  are determined  by the intersection points of a  logarithmic curve and a parabola. In order to find these points  we look at the numerator of the metric function:
 \begin{eqnarray}
n(r)= r^2-2Mr +\frac{\kappa^2q^2}{4}(1+5a_1)  +a_1\kappa^2q^2ln\frac{r}{r_0} =r^2f^2(r). \label{32}
\end{eqnarray}
The derivative of the function or $n\rq{}(r)$ has two local extrema at 
\begin{eqnarray}
r_{1}=    \frac{M}{2} + \frac{1}{2} \sqrt{M^2 -2a_1\kappa^2 q^2 }  \quad ,   \quad
r_{2}=    \frac{M}{2} - \frac{1}{2} \sqrt{M^2 -2a_1\kappa^2 q^2 } .
\end{eqnarray}
We  find the number of event horizons depending on the  critical values of the parameters $a_1 $ and $r_0$  by looking at the behavior of  the derivative of the above expression
 (\ref{32}) . 
Thus,  $n\rq{}(r)$  has  one root ($r_1$) for $ a_1 <  0   $,   two roots for $ 0  < a_1 <  \frac{M^2}{2•\kappa^2 q^2} $, one root for $a_1 = \frac{M^2}{2•\kappa^2 q^2}  $,  ($r_1=r_2$) and  no roots for $a_1 > \frac{M^2}{2•\kappa^2 q^2}  $.
The number  of the horizons will  differ  according to   the sign of the  function $n(r)$ 
at these critical points and the  values of   $a_1$ and  $r_0$. We will determine these intervals  and plot the graphs of the function $n(r)$ in the interval  for certain values of the parameters as follows.
\begin{figure}[h]{}
  \centering
       \includegraphics[width=0.4\textwidth]{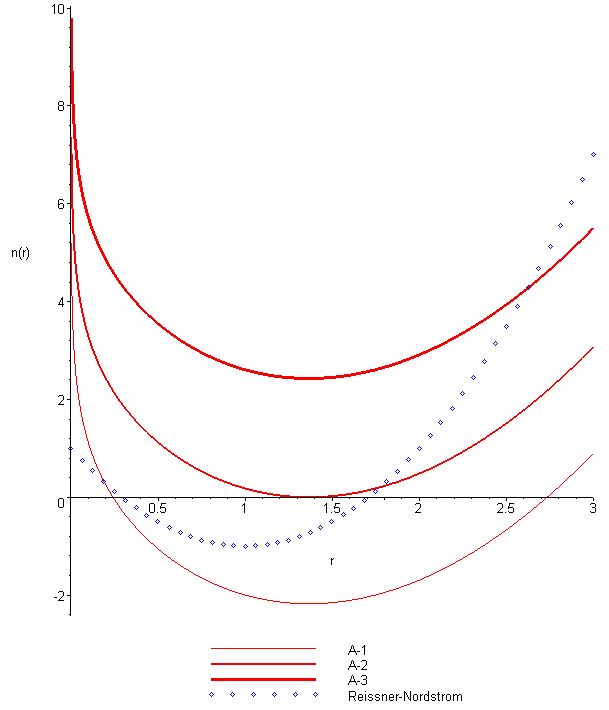} 
  \parbox{6.5in}{\small \caption{{{ \protect \small The graph of the  function $n(r)$    for the Case:A  with $a_1=-1$,  A-1) $ r_0 =1$, $  n(r_1)  \simeq -2.17 $,  A-2) $ r_0  \simeq 8.83 $, $ n(r_1)= 0$,    A-3)  $ r_0  = 100$,  $ n(r_1)\simeq  2.43$   and  Reisner-Nordstr\"{o}m case   is plotted by the dotted curve ($\kappa$ =M=q=1). }    }} }
  \end{figure}
  \begin{figure}[!h]{}
  \centering
    \subfloat[for $  a_1=0.25$   ]{ \includegraphics[width=0.4\textwidth]{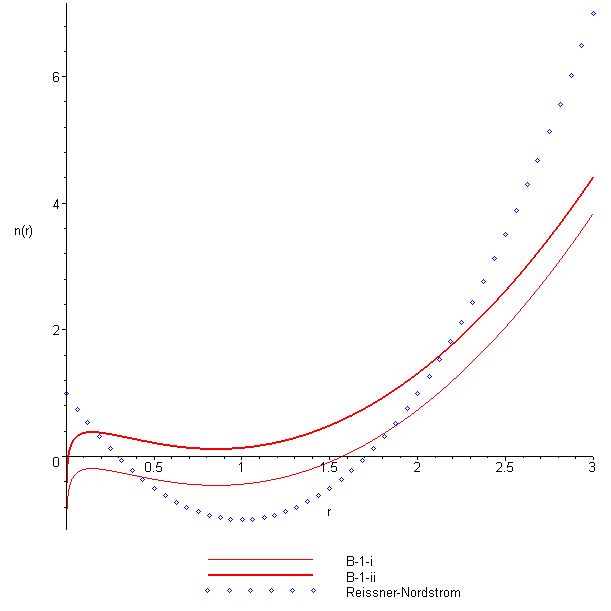}} 
        \subfloat[for $a_1=0.25 $ ]{ \includegraphics[width=0.4\textwidth]{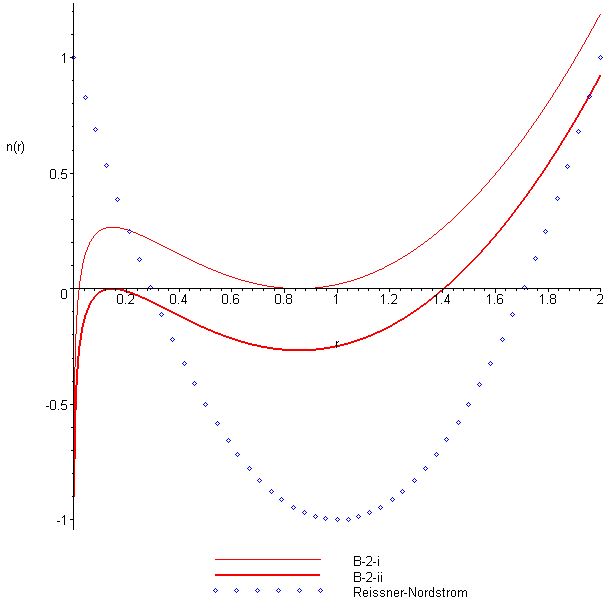}} \\
         \subfloat[for $a_1= 0.25, r_0=0.4  $]{ \includegraphics[width=0.4\textwidth]{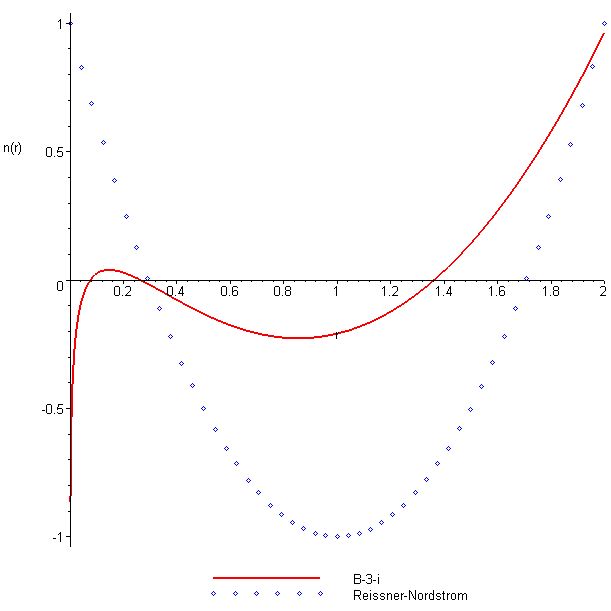}}
   {\small \caption{{{\small The graphics  of $n(r)$ for the Case-B, B-1-i)  $r_0=1$, $ n(r_1)\simeq - 0.45$, $n(r_2) \simeq -0.19  $,  B-1-ii) $r_0=0.1$,  $ n(r_1)\simeq 0.12$, $n(r_2) \simeq 0.38  $, 
   B-2-i) $  r_0\simeq 0.16$, \  $n(r_1)=0$,  $n(r_2) \simeq 0.26  $,  B-2-ii) $ r_0\simeq 0.47$, \  $n(r_1)\simeq - 0.26$, $n(r_2) = 0  $,  B-3-i)   $ n(r_1)\simeq - 0.22 $  and    $ n(r_2)= 0.04  $   and  Reisner-Nordstr\"{o}m case   is plotted by the dotted curve ($\kappa$ =M=q=1).}    }}   }
  \label{fig-f(r)}
\end{figure}

\begin{description}
\item[Case A] For $  a_1 < 0 $, we can determine the interval of $r_0$ related with  the numbers of horizons from the following inequalities (for a specific case see FIG.1 );
\begin{enumerate}
\item If $ n(r_1)< 0$,  there are  two horizons.
\item If $ n(r_1)=0$,  there is one horizon.
\item If $ n(r_1)>0$,  there is  no  horizon.
\end{enumerate}
\end{description}

  \newpage
\begin{description}
\item[Case B] For $0 < a_1 < \frac{M^2}{2•\kappa^2 q^2} $,  (see FIG.2);
\begin{enumerate}
\item If  i) $  n(r_1), n(r_2)<0$  or  ii) $ n(r_1), n(r_2)>0$ there is one horizon.
\item If i) $  n(r_1)=0$ and $n(r_2)>0 $  or  \ ii) $ n(r_1)< 0$ and $n(r_2)=0 $  there are two horizons.
\item  If  i) $ n(r_1)< 0$ and $n(r_2)>0 $  or  \  ii) $ n(r_1)>0$ and $n(r_2)< 0 $  there are three  horizons. But it is not possible to find an $r_0$ in the second interval $ n(r_1)>0$ and $n(r_2)< 0 $.
\end{enumerate}
\end{description}
\begin{figure}[h]{}
  \centering
              \includegraphics[width=0.4\textwidth]{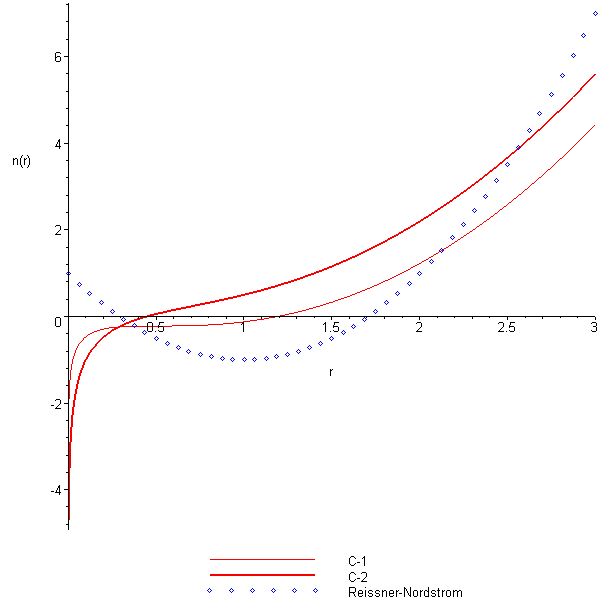}
  \parbox{6.5in}{  \caption{{{\small{The graph of  $n(r)$  C-1)  $a_1=0.5, r_0=1 $, $n(r_1)=n(r_2)\simeq -0.22 $,  C-2)  $a_1=1, r_0=1 $   and  Reisner-Nordstr\"{o}m case   is plotted by the dotted curve ($\kappa$ =M=q=1). }}     }}  }
  \label{fig-f(r)}
\end{figure}

\begin{description}
\item[Case C] For 1) $a_1  = \frac{M^2}{2•\kappa^2 q^2} $ and 2) $a_1  > \frac{M^2}{2•\kappa^2 q^2} $(see FIG.3);    $n(r)$  increases monotonically from $-\infty $    to  $\infty $    and there is one horizon.
\end{description}

The case with no horizon exhibits a  naked singularity. The cases with a single horizon resemble the extreme Reissner-Nordstr\"{o}m solution,  while the cases with two horizons correspond to the the generic  Reissner-Nordstr\"{o}m geometry.  The case with three horizons seems to be new.

\noindent Further investigation of the black hole properties can be done numerically for each of the cases above. We plan  to deal with this problem in a future work.


\newpage

\section{Conclusion}

\noindent We  considered a   non-minimally
$Y(R)F^2$-coupled   Einstein-Maxwell theory and looked for   static,
spherically symmetric solutions for a specific function $Y(R)  =\frac{1}{1-a_1ln{\frac{R}{R_0}}}$  that  shows charge  screening effects.  We obtained a class of asymptotically flat solutions that include  new black hole candidate configurations, except for the parameter values $a_1<0 $ and $n(r_1)>0$  when there is a naked essential  singularity at the origin. These particular solutions may shed light on some problems of gravity such as dark matter and dark energy without introducing a cosmological constant or any other exotic fields.
This means that, if dark matter is not some strange matter, but, for instance the non-minimal couplings produce such effects \cite{nojiri4};
then the electromagnetic potentials get modified at large (astrophysical) scales and thus contribute to the conventional electromagnetic energy density 
which may then be interpreted as the effects of dark matter.
 We also note that the electric charge q need not have  a large value to have observable  effects. If q is small, $a_1$ may be large so that the product $a_1q^2$ becomes important at large (astrophysical) scales. This can explain the rotation curves of galaxies for certain parameter values \cite{nojiri4}.

\vskip 1cm

\section{Acknowledgement}

\noindent
T.D. gratefully acknowledges partial  support from The
Turkish Academy of Sciences (TUBA).
The research of \"{O}.S. is
supported in part by a grant from The Scientific and Technological 
Research Council  of Turkey (TUBITAK).


\vskip 1cm


\end{document}